\documentclass[12pt,a4paper]{article}
\usepackage[latin1]{inputenc}
\usepackage{amsmath}
\usepackage{amsfonts}
\usepackage{amssymb}
\usepackage{graphicx}
\usepackage{authblk}
\begin{document}
	\title{Nonlinear supercoherent states and geometric phases for the supersymmetric harmonic oscillator}
	\author{Erik D\'iaz-Bautista 
		and David J. Fern\'andez C}

	\affil{\small Physics Department, Cinvestav, P.O. Box 14-740, 07000 Mexico City, Mexico}
	\date{\today}

	\maketitle
		\begin{abstract}
Nonlinear supercoherent states, which are eigenstates of nonlinear deformations of the Kornbluth-Zypman annihilation operator for the supersymmetric harmonic oscillator, will be studied. They turn out to be expressed in terms of nonlinear coherent states, associated to the corresponding deformations of the standard annihilation operator. We will discuss as well the Heisenberg uncertainty relation for a special particular case, in order to compare our results with those obtained for the Kornbluth-Zypman linear supercoherent states. As the supersymmetric harmonic oscillator executes an evolution loop, such that the evolution operator becomes the identity at a certain time, thus the linear and nonlinear supercoherent states turn out to be cyclic and the corresponding geometric phases will be evaluated.
\end{abstract}
	\section{\label{sec:level1}Introduction}
	In quantum field theory, supersymmetry (SUSY) generates transformations between boson and fermion, supplying us with a global framework to describe both kinds of particles using the same supersymmetric multiplet \cite{wess}. Witten was the first people who considered the simplest supersymmetric system composed of a particle of spin $1/2$ moving in one dimension \cite{witten1}, defining thus the algebra that the charge operators (also called supercharges) must meet and the way in which the supersymmetric Hamiltonian is expressed in terms of them. These algebraic relations define what nowadays is called supersymmetric quantum mechanic (SUSY QM).
	
	One of the simplest systems realizing SUSY QM is the so-called supersymmetric harmonic oscillator. It has been employed to illustrate how supersymmetry works, and it has been used also to introduce the supercoherent states, a supersymmetric generalization of the standard coherent states \cite{aragone}.
	
	In order to perform the supercoherent states analysis, it is important to study the algebraic commutation relations between the Hamiltonian $\hat{H}$ and the creation $\hat{A}^\dagger$ and annihilation operators $\hat{A}$ of the system. In fact, one starts by defining the SUSY annihilation operators (SAO) by imposing a natural commutation relationship with $\hat{H}$. However, as it was shown in \cite{aragone,hussin}, the form of these operators is not unique. In a recent work, Kornbluth and Zypman \cite{zypman}  have found a very general expression for the operator $\hat{A}$, as well as the explicit form for its eigenvectors $|Z\rangle$ with complex eigenvalues $z$, the so-called supercoherent states. The last turn out to be expressed in terms of standard coherent states \cite{glauber,glauber63,glauber631,glauber1}, defined for the one-dimensional harmonic oscillator through the Heisenberg-Weyl algebra.
	
	It is possible to consider also a class of deformations preserving the structure given in \cite{zypman}. In this work we will analyze a particular deformation of the SAO, the corresponding operator being labeled as $\hat{A}'$. Proceeding in a similar way as Kornbluth and Zypman, we are going to find the explicit form of its eigenvector $|\mathcal{Z}\rangle$ with complex eigenvalue $Z$. Due to the deformation employed, we will see that the eigenstates of $\hat{A}'$ are expressed now in terms of the corresponding nonlinear coherent states (NLCS), which have been studied previously for nonlinear deformations of Lie algebras \cite{ya,vinet,junker,junker99,karassiov}. Therefore, the states $|\mathcal{Z}\rangle$ will be called nonlinear supercoherent states (NLSCS).
	
	On the other hand, due to the equidistant nature of its Hamiltonian spectrum, the supersymmetric harmonic oscillator executes an evolution loop, a closed dynamical process such that the evolution operator becomes the identity (up to a global phase factor) at a certain time (the so-called loop period, $\tau = 2\pi/\omega$ in our case). The evolution loops were introduced as the basis to produce arbitrary unitary operations, which is the subject of the dynamical manipulation problem \cite{mielnik77,mielnik86,mielnik086,fernandez94}, and they appear for systems ruled either by time-dependent or by time-independent Hamiltonians \cite{mielnik77,mielnik86,mielnik086,fernandez94,djf92,djf94,djf12,wolf15}.
	
	It is important to note that, for systems executing an evolution loop, any quantum state turns out to be cyclic. Starting from the work of Berry \cite{berry84}, it has been realized that to any cyclic state one can associate a geometric phase $\beta$, which characterizes global curvature effects of the space of physical states. In fact, the geometric phase turns out to be the holonomy of the horizontal lifting of the closed trajectory in projective Hilbert space \cite{berry84,aharonov87,layton90,moore90,moore91,fernandez92a,fernandez92b,fernandez93,seleznyova93,solem93,campos93} (for a recent collection of articles about geometric phases see \cite{dennis10}). As a consequence, the linear and nonlinear supercoherent states of the supersymmetric harmonic oscillator become cyclic states, and it would be important to evaluate their associated geometric phases.
	
	This paper is organized  as follows. In section \ref{sec:level2} the standard and the simplest nonlinear coherent states will be quickly reviewed. In section \ref{sec:level3} the Hamiltonian of the supersymmetric harmonic oscillator, its eigenstates, eigenvalues, the annihilation and creation operators recently introduced by Kornbluth and Zipman, and its evolution loop will be discussed. In section \ref{sec:level4} a simple deformation of the previous ladder operators and the associated nonlinear supercoherent states will be constructed, while in section \ref{sec:level5} the corresponding geometric phases will be studied. Our conclusions will be presented in section \ref{sec:level6}.
	
	\section{\label{sec:level2}Nonlinear coherent states}
	
	Before addressing the nonlinear coherent states, let us quickly review the standard harmonic oscillator ones.
	\subsection{\label{subsec:level21}Standard coherent states}
	A standard harmonic oscillator coherent state $|\alpha\rangle$ can be defined by the relation
	\begin{equation}\label{12}
		\hat{a}|\alpha\rangle=\alpha|\alpha\rangle, \quad \alpha\in \mathbb{C}, \quad \langle\alpha|\alpha\rangle=1,
	\end{equation}
	\textit{i.e.}, it is a normalized eigenstate of the annihilation operator $\hat{a}$ with complex eigenvalue $\alpha$.
	
	In addition, other definitions can be used to build this type of quantum states, which for the harmonic oscillator are equivalent to each other:
	\begin{itemize}
		\item They satisfy the minimum Heisenberg uncertainty relation, 
		$(\sigma_x)_{\alpha}(\sigma_p)_{\alpha}=\hbar/2$, where $\sigma_x$ and $\sigma_p$ are the uncertainties associated with the position and momentum operators, respectively. 
		\item They appear of applying the unitary displacement operator $D(\alpha)=\exp(\alpha\hat{a}^{\dagger}-\alpha^\ast\hat{a})$ on the ground state, $|\alpha\rangle=D(\alpha)|0\rangle$; hereafter $z^\ast$ will denote the complex conjugate of $z\in \mathbb{C}$.
	\end{itemize}
	
	In the Fock representation, the normalized coherent states acquire the form
	\begin{equation}\label{13}
		|\alpha\rangle=\exp\left(-\frac{1}{2}|\alpha|^2\right)\sum_{n=0}^{\infty}\frac{\alpha^n}{\sqrt{n!}}|n\rangle.
	\end{equation}
	
	From this explicit form, the mean values for the position $\hat{x}$ and momentum $\hat{p}$ operators as well as their squares become (for $\hbar=m=\omega=1$):
	\begin{eqnarray}\label{14.0}
		\langle\hat{x}\rangle &=& \sqrt{2}\mathbf{Re}(\alpha), \\
		\langle\hat{x}^2\rangle &=& \frac{1}{2}+2[\mathbf{Re}(\alpha)]^2, \\
		\langle\hat{p}\rangle &=& \sqrt{2}\mathbf{Im}(\alpha), \\
		\langle\hat{p}^2\rangle &=& \frac{1}{2}+2[\mathbf{Im}(\alpha)]^2.
	\end{eqnarray}
	Hence, the Heisenberg uncertainty relation turns out to be given by $(\sigma_x)_{\alpha}(\sigma_p)_{\alpha}=1/2$.
	
	\subsection{\label{subsec:level22}Nonlinear coherent states}
	The nonlinear coherent states (NLCS) have attracted a lot of interest because they exhibit nonclassical behavior. Besides, many states in quantum optics can be viewed as a type of NLCS. Some authors \cite{manko,nlcs1,dodonov,nlcs} define a nonlinear coherent state $|z\rangle_{f}$ as an eigenstate of the deformed annihilation operator $\tilde{a}=f(\hat{N})\hat{a}$, so that
	\begin{equation}\label{15}
		\tilde{a}|z\rangle_{f}=f(\hat{N})\hat{a}|z\rangle_{f}=z|z\rangle_{f},
	\end{equation}
	where $f(\hat{N})$ is a well behaved real function of the number operator $\hat{N}=\hat{a}^{\dagger}\hat{a}$ and $z$ is the complex eigenvalue.
	
	Note that these deformations give place to a kind of anharmonic oscillators, called nonlinear $f$-oscillators, whose frequency depends on the energy \cite{manko}. Moreover, a classification of potentials with energy spectra equivalent to $f$-oscillators with different kinds of nonlinearity has been also performed \cite{mcd04}.
	
	Let us remark that the nonlinear coherent states $|z\rangle_f$ depend strongly on the action of the operator $\tilde{a}$ on the Fock states $|n\rangle$, which in turn depend on the explicit form of the function $f(\hat{N})$, namely,
	\begin{equation}\label{15.0}
		\tilde{a}|n\rangle=\sqrt{n}f(n-1)|n-1\rangle, \quad n=0,1,2,\ldots
	\end{equation}
	
	In order to make this fact clear, let us suppose first that $f(n)\neq0$, $\forall$ $n=0,1,2,\ldots$ An explicit calculation, by defining $g(n)\equiv f(n-1)$, leads to
	\begin{equation}\label{15.01}
		|z\rangle_{f}=\left[\sum_{n=0}^\infty\frac{|z|^{2n}}{n!\{[g(n)]!\}^2}\right]^{-1/2}\sum_{n=0}^{\infty}\frac{z^n}{\sqrt{n!}[g(n)]!}|n\rangle,
	\end{equation}
	where
	\begin{equation*}
		[g(n)]! = \begin{cases} 1 &\mbox{for } n \equiv 0 \\
			g(n)\cdots g(1) & \mbox{for } n > 1 \end{cases}
	\end{equation*}
	
	On the other hand, if $f(0)=0$ and $f(n)\neq0$ for $n>0$, then we get:
	\begin{equation}\label{15.02}
		|z\rangle_{f}=\left[\sum_{n=0}^\infty\frac{|z|^{2n}}{(n+1)!\{[f(n)]!\}^2}\right]^{-1/2}\sum_{n=0}^{\infty}\frac{z^n}{\sqrt{(n+1)!}[f(n)]!}|n+1\rangle.
	\end{equation}
	
	Next, let us analyze in more detail two particular forms of $f(\hat{N})$, leading us to two kinds of coherent states which are quite similar in form (see eqs. (\ref{15.01}) and (\ref{15.02})) although their physical properties are really different.
	
	\subsubsection{\label{subsec:level23}Nonlinear coherent states for $f(\hat{N})=\hat{N}+\hat{1}$.}
	First of all, let us assume that $f(\hat{N})=\hat{N}+\hat{1}$ in eq. (\ref{15}) so that $f(n)=n+1\neq0$ $\forall$ $n=0,1,2,\ldots$. Therefore, the corresponding eigenvalue equation takes the form
	\begin{equation}\label{22}
		\tilde{a}|\alpha\rangle_{\mathrm{nl}}=(\hat{N}+\hat{1})\hat{a}|\alpha\rangle_{\mathrm{nl}}=\alpha|\alpha\rangle_{\mathrm{nl}}, \quad \alpha\in \mathbb{C},
	\end{equation}
	\textit{i.e.}, $|\alpha\rangle_{\mathrm{nl}}$ is an eigenvector of the operator $\tilde{a}=(\hat{N}+\hat{1})\hat{a}$ with eigenvalue $\alpha$.
	
	Thus, the explicit expression for the nonlinear coherent states of eq. (\ref{15.01}) becomes now:
	\begin{equation}\label{15.03}
		|\alpha\rangle_{\mathrm{nl}}=[_0F_2(1,1;r^2)]^{-1/2}\sum_{n=0}^\infty\frac{\alpha^n}{\Gamma(n+1)\sqrt{\Gamma(n+1)}}|n\rangle,
	\end{equation}
	where $r=|\alpha|$ and $_pF_q$ is the generalized hypergeometric function defined by
	\begin{equation}\label{24.1}  _pF_q(a_1,\ldots,a_p,b_1,\ldots,b_q;x)=\frac{\Gamma(b_1)\ldots\Gamma(b_q)}{\Gamma(a_1)\ldots\Gamma(a_p)}\sum_{n=0}^{\infty}\frac{\Gamma(a_1+n)\ldots\Gamma(a_p+n)}{\Gamma(b_1+n)\ldots\Gamma(b_q+n)}\frac{x^n}{n!}.
	\end{equation}
	
	Knowing the explicit form of the nonlinear coherent states $|\alpha\rangle_{\mathrm{nl}}$, the mean values of the operators $\hat{x}$, $\hat{p}$ and their squares become:
	\begin{eqnarray}\label{24.2}
		\langle\hat{x}\rangle_{\mathrm{nl}} &=& \sqrt{2}\mathbf{Re}(\alpha)\frac{_0F_2(1,2;r^2)}{_0F_2(1,1;r^2)}, \label{11a}\\
		\langle\hat{x}^2\rangle_{\mathrm{nl}} &=& \frac{1}{2}+[\mathbf{Re}(\alpha)]^2\beta(r)+[\mathbf{Im}(\alpha)]^2\sigma(r), \label{11b}\\
		\langle\hat{p}\rangle_{\mathrm{nl}} &=& \sqrt{2}\mathbf{Im}(\alpha)\frac{_0F_2(1,2;r^2)}{_0F_2(1,1;r^2)}, \label{11c}\\
		\langle\hat{p}^2\rangle_{\mathrm{nl}} &=& \frac{1}{2}+[\mathbf{Re}(\alpha)]^2\sigma(r)+[\mathbf{Im}(\alpha)]^2\beta(r), \label{11d}
	\end{eqnarray}
	where
	\begin{eqnarray*}\label{24.3}
		\beta(r)&=&\frac{_0F_2(2,2;r^2)}{_0F_2(1,1;r^2)}+\frac{1}{2}\frac{_0F_2(1,3;r^2)}{_0F_2(1,1;r^2)}, \nonumber\\ \sigma(r)&=&\frac{_0F_2(2,2;r^2)}{_0F_2(1,1;r^2)}-\frac{1}{2}\frac{_0F_2(1,3;r^2)}{_0F_2(1,1;r^2)}.
	\end{eqnarray*}
	
	\begin{figure*}
		\centering
		\includegraphics[width=14pc]{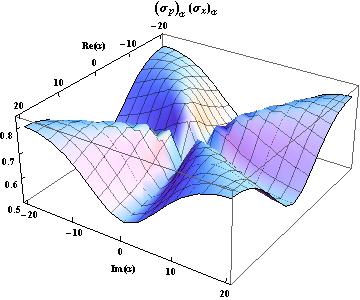} \hspace{2pc}
		\includegraphics[width=10pc]{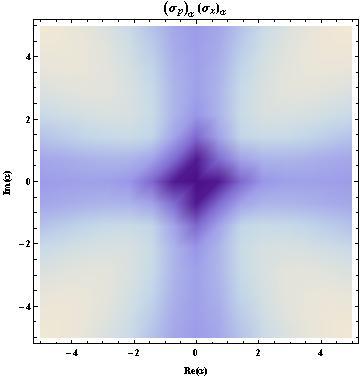}
		\caption{\label{fig:nl}Uncertainty relation $(\sigma_x)_\alpha(\sigma_p)_\alpha$ as function of $\alpha$ (left) for the nonlinear coherent states of eq. (\ref{15.03}). The corresponding uncertainty density is shown on the right.}
	\end{figure*}
	
	Then, the explicit expression for the Heisenberg uncertainty relation is (see its plot in fig. \ref{fig:nl}):
	\small
	\begin{equation}\label{26}
		(\sigma_x)
		_\alpha(\sigma_p)
		_\alpha=\sqrt{\frac{1}{2}-[\mathbf{Re}(\alpha)]^2\tau(r)+[\mathbf{Im}(\alpha)]^2\sigma(r)}\sqrt{\frac{1}{2}-[\mathbf{Im}(\alpha)]^2\tau(r)+[\mathbf{Re}(\alpha)]^2\sigma(r)},
	\end{equation}
	\normalsize
	where
	\begin{equation*}\label{27}
		\tau(r)=2\left[\frac{_0F_2(1,2;r^2)}{_0F_2(1,1;r^2)}\right]^2-\beta(r).
	\end{equation*}
	
	Note that in the limit $\alpha\rightarrow0$, the uncertainty relation takes a minimum value equal to $1/2$, which is consistent with the fact that $\lim\limits_{\alpha\rightarrow0}|\alpha\rangle_{\mathrm{nl}}=|0\rangle$.
	
	\subsubsection{\label{subsec:level24}Nonlinear coherent states for $f(\hat{N})=\hat{N}$.}
	Now, let us assume that $f(\hat{N})=\hat{N}$ in eq. (\ref{15}), so that $f(0)=0$. Similarly as in the previous section, if $|\alpha\rangle_{\mathrm{NL}}$ label the eigenvectors of $\tilde{a}=\hat{N}\hat{a}$, the eigenvalue equation becomes
	\begin{equation}\label{22.1}
		\tilde{a}|\alpha\rangle_{\mathrm{NL}}=\hat{N}\hat{a}|\alpha\rangle_{\mathrm{NL}}=\alpha|\alpha\rangle_{\mathrm{NL}}, \quad \alpha\in \mathbb{C}.
	\end{equation}
	Thus, the nonlinear coherent states of eq.~(\ref{15.02}) are expressed as
	\begin{equation}\label{23}
		|\alpha\rangle_{\mathrm{NL}}=\left[_0F_2(1,2;r^2)\right]^{-1/2}\sum_{n=0}^\infty\frac{\alpha^{n}}{\Gamma(n+1)\sqrt{\Gamma(n+2)}}|n+1\rangle.
	\end{equation}
	
	Once again, the mean values of the operators $\hat{x}$, $\hat{p}$ and their squares in the state $|\alpha\rangle_{\mathrm{NL}}$ are:
	\begin{eqnarray}\label{25}
		\langle\hat{x}\rangle_{\mathrm{NL}} &=& \sqrt{2}\mathbf{Re}(\alpha)\frac{_0F_2(2,2;r^2)}{_0F_2(1,2;r^2)}, \label{17a}\\
		\langle\hat{x}^2\rangle_{\mathrm{NL}} &=& \frac{3}{2}+[\mathbf{Re}(\alpha)]^2\,\frac{_0F_2(2,3;r^2)}{_0F_2(1,2;r^2)}, \label{17b} \\
		\langle\hat{p}\rangle_{\mathrm{NL}} &=& \sqrt{2}\mathbf{Im}(\alpha)\frac{_0F_2(2,2;r^2)}{_0F_2(1,2;r^2)}, \label{17c}\\
		\langle\hat{p}^2\rangle_{\mathrm{NL}} &=& \frac{3}{2}+[\mathbf{Im}(\alpha)]^2\,\frac{_0F_2(2,3;r^2)}{_0F_2(1,2;r^2)}. \label{17d}
	\end{eqnarray}
	\begin{figure*}
		\centering
		\includegraphics[width=14pc]{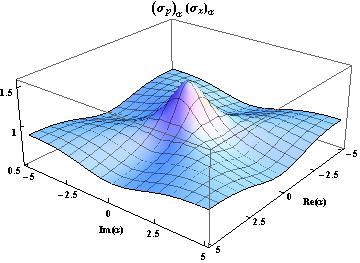} \hspace{2pc}
		\includegraphics[width=9pc]{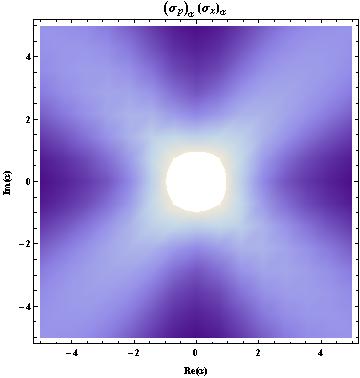}
		\caption{\label{fig:nl1}Uncertainty relation $(\sigma_x)_\alpha(\sigma_p)_\alpha$ as function of $\alpha$ (left) for the nonlinear coherent states of eq. (\ref{23}). The corresponding uncertainty density is shown on the right.}
	\end{figure*}
	Then, the expression for the Heisenberg uncertainty relation is now  \cite{fhn94} (see also fig. \ref{fig:nl1}):
	\begin{equation}\label{26.1}
		(\sigma_x)
		_\alpha(\sigma_p)
		_\alpha=\sqrt{\frac{3}{2}-[\mathbf{Re}(\alpha)]^2\rho(r)}\sqrt{\frac{3}{2}-[\mathbf{Im}(\alpha)]^2\rho(r)},
	\end{equation}
	where
	\begin{equation*}\label{27.1}
		\rho(r)=2\left[\frac{_0F_2(2,2;r^2)}{_0F_2(1,2;r^2)}\right]^2-\frac{_0F_2(2,3;r^2)}{_0F_2(1,2;r^2)}.
	\end{equation*}
	
	In the limit $\alpha\rightarrow0$, the uncertainty relation takes now a maximum value equal to $3/2$, which is also consistent with the fact that $\lim\limits_{\alpha\rightarrow0}|\alpha\rangle_{\mathrm{NL}}=|1\rangle$.
	
	\section{\label{sec:level3}Supersymmetric harmonic oscillator}\label{SUSY}
	The supercoherent states are defined as eigenstates of the annihilation operator of the supersymmetric harmonic oscillator which Hamiltonian  $\hat{H}$, expressed in terms of the standard creation and annihilation operators with  $\hbar=m=1$, is \cite{aragone}:
	\begin{equation}\label{29}
		\hat{H}=\omega\left(
		\begin{array}{cc}
			\hat{a}^{\dagger}\hat{a} & 0 \\
			0 & \hat{a}\hat{a}^{\dagger} \\
		\end{array}
		\right).
	\end{equation}
	
	Its corresponding energy levels and associated eigenstates are:
	\begin{eqnarray}\label{30}
		E_{n} \! = \! n\omega, & \quad \vert\Psi_n^+\rangle \! = \! \left(
		\begin{array}{c}
			|n\rangle \\
			0 \\
		\end{array}
		\right) \! , \quad
		\vert\Psi_n^-\rangle \! = \!
		\left(
		\begin{array}{c}
			0 \\
			|n-1\rangle \\
		\end{array}
		\right) \! ,
	\end{eqnarray}
	where $n=0,1,2,\dots$ and $\vert\Psi_0^-\rangle\equiv 0$. Note that the eigenspace associated to $E_n$ for $n\neq0$ is two-dimensional, \textit{i.e.}, $E_n$ is doubly degenerate
	for such an $n$.
	
	\subsection{\label{subsec:level31}Supersymmetric annihilation operator (SAO)}
	By definition, a supersymmetric annihilation operator $\hat{X}$ satisfies the following commutation relation with $\hat{H}$
	\begin{equation}\label{30.1}
		[\hat{H},\hat{X}]=-\omega \hat{X}.
	\end{equation}
	
	As discussed in \cite{hussin}, the SAO associated to the Hamiltonian $\hat{H}$ of the SUSY harmonic oscillator is not unique. A very general form, which includes most of the previous proposals, has been considered recently \cite{zypman}, namely,
	\begin{equation}\label{39}
		\hat{A}=\left(
		\begin{array}{cc}
			k_1\hat{a} & k_2 \\
			k_3\hat{a}^2 & k_4\hat{a} \\
		\end{array}
		\right),
	\end{equation}
	where $k_i$ are arbitrary complex numbers.
	
	For some particular values of the parameters $k_i$, the annihilation operators given in \cite{aragone} are recovered (see also \cite{hussin}). Indeed, for $k_1=k_2=k_4=1$ and $k_3=0$ the corresponding annihilation operator becomes
	\begin{equation}\label{39.1}
		\hat{A}=\left(
		\begin{array}{cc}
			\hat{a} & 1 \\
			0 & \hat{a} \\
		\end{array}
		\right),
	\end{equation}
	while for $k_1=k_4=1$ and $k_2=k_3=0$, it takes the form
	\begin{equation}\label{39.2}
		\hat{A}=\left(
		\begin{array}{cc}
			\hat{a} & 0 \\
			0 & \hat{a} \\
		\end{array}
		\right).
	\end{equation}
	
	The supercoherent states are defined in the usual way
	\begin{equation}\label{40}
		\hat{A}|Z\rangle=z|Z\rangle,
	\end{equation}
	which are expanded in terms of the energy eigenstates of the SUSY harmonic oscillator in the form
	\begin{equation}\label{40.1}
		|Z\rangle = \sum\limits_{n=0}^{\infty} a_n \vert\Psi_n^+\rangle + \sum\limits_{n=1}^{\infty} c_n \vert\Psi_n^-\rangle .
	\end{equation}
	
	They are going to be expressed in terms of the eigenvalues $\kappa_{\pm}$ of the matrix $K$, whose entries are given by the parameters $k_i$ as
	\begin{equation}\label{40.41}
		K\equiv\left(
		\begin{array}{cc}
			k_1 & k_2 \\
			k_3 & k_4 \\
		\end{array}
		\right).
	\end{equation}
	
	This induces a classification of the supercoherent states $|Z\rangle$ into three different families: degenerate ($\kappa_+=\kappa_-\neq0$); singular ($\kappa_+\kappa_-=0$); generic (anything else). The evaluation of the supercoherent states of eq.~(\ref{40}) for each of these three cases has been performed in \cite{zypman}. Since this calculation is similar for our deformed SAO, we will make it later on, when we will introduce the deformation addressed in this paper. Before doing this, let us discuss first the evolution loop of the system and the geometric phase associated to an arbitrary cyclic state.
	
	\subsection{\label{subsec:level32}Evolution loop and geometric phase}
	The dynamics of a quantum system is determined by its evolution operator, a unitary operator $\hat{U}(t)$ which satisfies
	\begin{equation}\label{75}
		\frac{d\hat{U}(t)}{dt}=-i\hat{H}(t)\hat{U}(t),  \quad \hat{U}(0)=\hat{I},
	\end{equation}
	where the hermitian operator $\hat{H}(t)$ is the system Hamiltonian, $\hat{I}$ is the identity operator. In general, if at $t=0$ the system is
	in the state $\vert\Psi\rangle$, then at time $t\geq 0$ the system will evolve into $\vert\Psi(t)\rangle =
	\hat{U}(t)\vert\Psi\rangle$.
	
	An evolution loop (EL) is a circular dynamical processes such that $\hat{U}(t)$ becomes the identity
	(up to a phase factor) at a certain time,
	\begin{equation}\label{76}
		\hat{U}(\tau)=\exp(i\Omega)\hat{I},
	\end{equation}
	where $\tau>0$ is the loop period and $\Omega\in {\mathbb R}$ \cite{mielnik77,mielnik86,mielnik086,djf94}. It is important to
	realize that, for a system executing such a particular evolution, any state $\vert\Psi\rangle$ turns out to be cyclic,
	$$
	\vert\Psi(\tau)\rangle = \hat{U}(\tau)\vert\Psi\rangle = \exp(i\Omega) \vert\Psi\rangle.
	$$	
	
	On the other hand, it has been realized that to any cyclic state such that $\vert\Psi(\tau)\rangle = \exp(i\Omega) \vert\Psi\rangle$
	one can associate a geometric phase, which is given by
	\begin{equation}\label{77}
		\beta=\Omega+\int_0^\tau \langle\Psi(t)\vert \hat{H}(t)\vert\Psi(t)\rangle dt.
	\end{equation}
	In particular, if the cyclic evolution is induced by a time-independent Hamiltonian $\hat{H}$, the geometric phase becomes
	\begin{equation}\label{gptih}
		\beta=\Omega+\tau\langle\Psi|\hat{H}|\Psi\rangle.
	\end{equation}
	
	Let us note that, since the spectrum of the SUSY harmonic oscillator consists of equidistant energy levels (compare eq.~\ref{30}), then it executes an evolution loop of period $\tau=2\pi/\omega$. In fact, it turns out that that \cite{djf94}:
	\begin{eqnarray}
		\hat{U}(\tau) & = & \exp(-iH\tau) \sum_{n=0}^\infty \sum_{\epsilon=+,-} \vert \Psi_n^\epsilon\rangle \langle \Psi_n^\epsilon \vert \nonumber \\
		& = & \sum_{n=0}^\infty \sum_{\epsilon=+,-} \exp(-in\omega\tau) \vert \Psi_n^\epsilon\rangle \langle \Psi_n^\epsilon \vert = \hat{I}.
	\end{eqnarray}
	Thus, any state $\vert\Psi\rangle$ of the SUSY harmonic oscillator turns out to be cyclic and it has associated a geometric phase which can be
	calculated using eq. (\ref{gptih}) with $\Omega = 0$ (mod$(2\pi)$) and $\tau=2\pi/\omega$.
	
	Let us discuss next the nonlinear deformations of the supersymmetric annihilation operator we are interested in and their corresponding supercoherent states.
	
	\section{\label{sec:level4}Nonlinear supercoherent states}
	The lack of uniqueness of the supersymmetric annihilator operator is clear from the results discussed in the previous section (see also \cite{hussin,zypman}); thus, we introduce next a deformation of this operator, obtain then their eigenstates and compare them with the supercoherent states of \cite{zypman}. We could consider a generic operator of the form $\tilde{a}=f(\hat{N})\hat{a}$ \cite{nlcs}; however, to simplify the calculations we will choose the simplest cases, for $f(\hat{N})=\hat{N}+\hat{1}$ and $f(\hat{N})=\hat{N}$ (see eqs.~(\ref{22}) and (\ref{22.1})).
	
	\subsection{\label{subsec:level41}First deformation of the SAO}
	Let us make the substitution $\hat{a}\rightarrow(\hat{N}+\hat{1})\hat{a}$ in eq. (\ref{39}) so that the deformed annihilation operator $\hat{A}'$ becomes
	\begin{equation}\label{44}
		\hat{A}'=\left(
		\begin{array}{cc}
			k_1(\hat{N}+\hat{1})\hat{a} & k_2 \\
			k_3[(\hat{N}+\hat{1})\hat{a}]^2 & k_4(\hat{N}+\hat{1})\hat{a} \\
		\end{array}
		\right).
	\end{equation}
	
	Let $|\mathcal{Y}\rangle$ be the eigenstate of the operator $\hat{A}'$, such that
	\begin{equation}\label{45}
		\hat{A}'|\mathcal{Y}\rangle=Y|\mathcal{Y}\rangle, \quad |\mathcal{Y}\rangle=
		\sum\limits_{n=0}^{\infty} s_n \vert\Psi_n^+\rangle + \sum\limits_{n=1}^{\infty} t_n \vert\Psi_n^-\rangle .
	\end{equation}
	
	Using this eigenvalue equation, the following recurrence relations are obtained
	\begin{equation}\label{46}
		\left\{
		\begin{array}{c}
			k_1(n+1)\sqrt{n+1}s_{n+1}+k_2t_{n+1}=Ys_{n}, \\
			k_3n(n+1)\sqrt{n}\sqrt{n+1}s_{n+1}+k_4n\sqrt{n}t_{n+1}=Yt_n. \\
		\end{array}
		\right.
	\end{equation}
	From the first expression with $n=0$ 
	we obtain that
	\begin{equation}\label{47}
		k_1s_1=Ys_0-k_2t_1,
	\end{equation}
	\textit{i.e.}, $s_0$ and $t_1$ are free parameters.
	
	By defining the
	quantities $\tilde{s}_{n}=\sqrt{[n!]^3}Y^{1-n}s_n$ and $\tilde{t}_n=\sqrt{[(n-1)!]^3}Y^{1-n}t_n$ for $n\geq1$, the system (\ref{46}) can be expressed as
	\begin{equation}\label{48}
		K\left(
		\begin{array}{c}
			\tilde{s}_{n+1} \\
			\tilde{t}_{n+1} \\
		\end{array}
		\right)=\left(
		\begin{array}{c}
			\tilde{s}_{n} \\
			\tilde{t}_{n} \\
		\end{array}
		\right),
	\end{equation}
	where $K$ is the matrix defined in eq. (\ref{40.41}). We denote the eigenvalues of the matrix $K$ by $\kappa_{\pm}$ and, when they are nonzero, define the quantities $\varphi_{\pm}\equiv Y\kappa_{\pm}^{-1}$.
	
	Similar to the nondeformed case, due to the dependence of $\kappa_{\pm}$ on the parameters $k_i$, the nonlinear supercoherent states $|\mathcal{Y}\rangle$ can be grouped in three cases.
	
	\subsubsection{Classification of the deformed families}
	\paragraph{\textbf{Generic}.} The solution of eq. (\ref{48}) is
	\begin{equation}\label{49}
		\left(
		\begin{array}{c}
			s_{n+1} \\
			t_{n+1} \\
		\end{array}
		\right)=s_0\left(
		\begin{array}{c}
			\frac{Y}{k_1(n+1)!\sqrt{(n+1)!}}d_{11} \\
			\frac{Y}{k_1n!\sqrt{n!}}d_{21} \\
		\end{array}
		\right)+t_1\left(
		\begin{array}{c}
			\frac{1}{(n+1)!\sqrt{(n+1)!}}\left(d_{12}-d_{11}\frac{k_2}{k_1}\right) \\
			\frac{1}{n!\sqrt{n!}}\left(d_{22}-d_{21}\frac{k_2}{k_1}\right) \\
		\end{array}
		\right),
	\end{equation}
	where
	 \footnotesize
	\begin{equation*}\label{50}
		\left(
		\begin{array}{cc}
			d_{11} & d_{12} \\
			d_{21} & d_{22} \\
		\end{array}
		\right)=\frac{1}{\kappa_{+}-\kappa_{-}}\left(
		\begin{array}{cc}
			(\kappa_{+}-k_4)\varphi_{+}^n-(\kappa_{-}-k_4)\varphi_{-}^n & k_2(\varphi_{+}^n-\varphi_{-}^n) \\
			k_3(\varphi_{+}^n-\varphi_{-}^n) & (\kappa_{+}-k_4)\varphi_{-}^n-(\kappa_{-}-k_4)\varphi_{+}^n \\
		\end{array}
		\right)
	\end{equation*}
	 \normalsize
	
	The supercoherent states $|\mathcal{Y}\rangle$ thus take the form
	\begin{equation}\label{50.0}
		|\mathcal{Y}\rangle=s'_0|\mathcal{Y}_\mathrm{A}\rangle+t'_1|\mathcal{Y}_\mathrm{C}\rangle,
	\end{equation}
	where $s'_0=s_0k_1^{-1}$, $t'_1=t_1(Yk_1)^{-1}$ and
	\begin{eqnarray}\label{51}
		|\mathcal{Y}_\mathrm{A}\rangle&=&\frac{1}{\kappa_{+}-\kappa_{-}}G_\mathrm{A}^{\mathrm{nl}}\left(
		\begin{array}{c}
			|\varphi_{+}\rangle_{\mathrm{nl}} \\
			|\varphi_{-}\rangle_{\mathrm{nl}} \\
		\end{array}
		\right), \\ |\mathcal{Y}_\mathrm{C}\rangle&=&\frac{1}{\kappa_{+}-\kappa_{-}}G_\mathrm{C}^{\mathrm{nl}}\left(
		\begin{array}{c}
			|\varphi_{+}\rangle_{\mathrm{nl}} \\
			|\varphi_{-}\rangle_{\mathrm{nl}} \\
		\end{array}
		\right),
	\end{eqnarray}
	with
	\begin{eqnarray*}\label{52}
		G_\mathrm{A}^{\mathrm{nl}} &=& 
		\left(
		\begin{array}{cc}
			\kappa_{+}(\kappa_{+}-k_4) & -\kappa_{-}(\kappa_{-}-k_4) \\
			k_3Y & -k_3Y \\
		\end{array}
		\right),
		\\
		G_\mathrm{C}^{\mathrm{nl}} &=&  
		\left(
		\begin{array}{cc}
			k_2\kappa_{+}\kappa_{-} & -k_2\kappa_{+}\kappa_{-} \\
			Y[k_1\kappa_{+}-(k_1^2+k_2k_3)] & -Y[k_1\kappa_{-}-(k_1^2+k_2k_3)] \\
		\end{array}
		\right).
	\end{eqnarray*}
	
	Besides the set $\{|\mathcal{Y}_\mathrm{A}\rangle,|\mathcal{Y}_\mathrm{C}\rangle\}$, we can choose an alternative set of nonlinear supercoherent states given by
	\begin{equation}\label{52.4}
		|\mathcal{Y}_{\pm}\rangle_{\mathrm{nl}}=\left(
		\begin{array}{c}
			k_2\kappa_{\pm}|\varphi_{\pm}\rangle_{\mathrm{nl}} \\
			(\kappa_{\pm}-k_1)Y|\varphi_{\pm}\rangle_{\mathrm{nl}} \\
		\end{array}
		\right),
	\end{equation}
	whereby it is possible to pass from the parameter space $\{k_1,k_2,k_3,k_4\}$ to the new one, generated by $\{\kappa_{+},\kappa_{-},k_1,k_2\}$.
	
	\paragraph{\textbf{Degenerate}.} This family is obtained if and only if $(k_1-k_4)^2+4k_2k_3=0$. Explicitly, the deformed supercoherent states become now
	\begin{equation}\label{52.0}
		|\mathcal{Y}\rangle=s'_0|\mathcal{Y}^\mathrm{d}_\mathrm{A}\rangle_{\mathrm{nl}}+t'_1|\mathcal{Y}^\mathrm{d}_\mathrm{C}\rangle_{\mathrm{nl}},
	\end{equation}
	where
	\begin{eqnarray}\label{52.1}
		|\mathcal{Y}^\mathrm{d}_\mathrm{A}\rangle_{\mathrm{nl}}&=&G_\mathrm{A}^{d,\mathrm{nl}}\left(
		\begin{array}{c}
			|\varphi\rangle_{\mathrm{nl}} \\
			|\varphi'\rangle_{\mathrm{nl}} \\
		\end{array}
		\right), \\
		|\mathcal{Y}^\mathrm{d}_\mathrm{C}\rangle_{\mathrm{nl}}&=&G_\mathrm{C}^{d,\mathrm{nl}}\left(
		\begin{array}{c}
			|\varphi\rangle_{\mathrm{nl}} \\
			|\varphi'\rangle_{\mathrm{nl}} \\
		\end{array}
		\right),
	\end{eqnarray}
	with $|\varphi'\rangle=\frac{d}{d\varphi}|\varphi\rangle$ and
	\begin{equation*}\label{52.2}
		G_\mathrm{A}^{d,\mathrm{nl}} = 
		\left(
		\begin{array}{cc}
			k_1 & -(\kappa-k_4)\varphi \\
			0 & -k_3\varphi^2 \\
		\end{array}
		\right),
		\quad
		G_\mathrm{C}^{d,\mathrm{nl}} =  
		\kappa\left(
		\begin{array}{cc}
			0 & -k_2\varphi \\
			k_1\varphi & -(\frac{k_4-k_1}{2})\varphi^2 \\
		\end{array}
		\right).
	\end{equation*}
	
	\paragraph{\textbf{Singular}.} If the matrix $K$ is singular then $k_1k_4=k_2k_3$. Hence, the corresponding deformed supercoherent state acquires the form
	\begin{equation}\label{52.3}
		|\mathcal{Y}\rangle_{\mathrm{nl}}=\left(
		\begin{array}{c}
			k_1|\varphi\rangle_{\mathrm{nl}} \\
			k_3\varphi|\varphi\rangle_{\mathrm{nl}} \\
		\end{array}
		\right),
	\end{equation}
	with $\varphi\equiv Y(k_1+k_4)^{-1}$.
	
	\subsubsection{Superposition and deformed uncertainties.}
	Consider a linear combination of the states $|\mathcal{Y}_{\pm}\rangle_{\mathrm{nl}}$ of eq. (\ref{52.4}) with parameters $\eta$ and $\lambda$ as follows:
	\begin{eqnarray}\label{52.5}
		|\mathcal{Y}_m\rangle_{\mathrm{nl}} &=& \cos\eta|\mathcal{Y}_+\rangle_{\mathrm{nl}}+\exp(i\lambda)\sin\eta|\mathcal{Y}_-\rangle_{\mathrm{nl}}\nonumber\\
		&=& \left(
		\begin{array}{c}
			\gamma_{1+}|\varphi_+\rangle_{\mathrm{nl}}+\gamma_{1-}|\varphi_-\rangle_{\mathrm{nl}} \\
			\gamma_{2+}|\varphi_+\rangle_{\mathrm{nl}}+\gamma_{2-}|\varphi_-\rangle_{\mathrm{nl}} \\
		\end{array}
		\right),
	\end{eqnarray}
	where the quantities $\gamma_{1\pm}$ and $\gamma_{2\pm}$ are given by
	\begin{eqnarray}\label{52.55}
		\gamma_{1+}=&\hspace{-0.5cm}k_2\kappa_+\cos\eta, \quad &\gamma_{1-}=k_2\kappa_-\exp(i\lambda)\sin\eta, \label{45a}\\
		\gamma_{2+}=&\:(\kappa_+-k_1)\cos\eta, \quad &\gamma_{2-}=(\kappa_--k_1)\exp(i\lambda)\sin\eta. \label{45b}
	\end{eqnarray}
	
	To find the uncertainties of the position and momentum operators in the nonlinear supercoherent state of eq. (\ref{52.5}), we use the following expressions involving two nonlinear coherent states $|\alpha_1\rangle_{\mathrm{nl}}$ and $|\alpha_2\rangle_{\mathrm{nl}}$:
	\begin{eqnarray}\label{52.6}
		\langle\alpha_1|\alpha_2\rangle_{\mathrm{nl}} &=& \frac{_0F_2(1,1;\alpha^{\ast}_1\alpha_2)}{\sqrt{_0F_2(1,1;|\alpha_1|^2)}\sqrt{_0F_2(1,1;|\alpha_2|^2)}}
		,\\
		\langle\alpha_1|\hat{x}|\alpha_2\rangle_{\mathrm{nl}} &=& \frac{\alpha_2+\alpha^{\ast}_1}{\sqrt{2}}\frac{_0F_2(1,2;\alpha^{\ast}_1\alpha_2)}{\sqrt{_0F_2(1,1;|\alpha_1|^2)}\sqrt{_0F_2(1,1;|\alpha_2|^2)}}, \\
		\nonumber  \langle\alpha_1|\hat{x}^2|\alpha_2\rangle_{\mathrm{nl}} &=& \frac{1}{2\sqrt{_0F_2(1,1;|\alpha_1|^2)}\sqrt{_0F_2(1,1;|\alpha_2|^2)}} \\
		&& \times \bigg[\frac{(\alpha_2^2+\alpha^{\ast2}_1)}{2}\,_0F_2(1,3;\alpha^{\ast}_1\alpha_2)+2\alpha_1^{\ast}\alpha_2\nonumber\\
		&& \times\:_0F_2(2,2;\alpha^{\ast}_1\alpha_2)+\:_0F_2(1,1;\alpha^{\ast}_1\alpha_2)\bigg], \\
		\langle\alpha_1|\hat{p}\:|\alpha_2\rangle_{\mathrm{nl}} &=& \frac{\alpha_2-\alpha^{\ast}_1}{\sqrt{2}\:i}\frac{_0F_2(1,2;\alpha^{\ast}_1\alpha_2)}{\sqrt{_0F_2(1,1;|\alpha_1|^2)}\sqrt{_0F_2(1,1;|\alpha_2|^2)}}, \\
		\nonumber  \langle\alpha_1|\hat{p}^2|\alpha_2\rangle_{\mathrm{nl}} &=& \frac{1}{2\sqrt{_0F_2(1,1;|\alpha_1|^2)}\sqrt{_0F_2(1,1;|\alpha_2|^2)}} \\
		&& \times \bigg[-\frac{(\alpha_2^2+\alpha^{\ast2}_1)}{2}\,_0F_2(1,3;\alpha^{\ast}_1\alpha_2)+2\alpha_1^{\ast}\alpha_2\nonumber\\
		&& \times\:_0F_2(2,2;\alpha^{\ast}_1\alpha_2)+\:_0F_2(1,1;\alpha^{\ast}_1\alpha_2)\bigg].
	\end{eqnarray}
	Note that, when $|\alpha_1\rangle_{\mathrm{nl}}=|\alpha_2\rangle_{\mathrm{nl}}\equiv|\alpha\rangle_{\mathrm{nl}}$ they reduce to those shown in eqs. (\ref{11a})-(\ref{11d}).
	
	The uncertainties we are looking for become now:
\footnotesize
	\begin{eqnarray}\label{52.7}
		\nonumber  \sigma^{2 (\mathrm{nl})}_x &=&\Delta_1^{-1}\left\{\Gamma^{+}\left[[\mathbf{Re}(\varphi_{+})]^2\beta(|\varphi_+|)+[\mathbf{Im}(\varphi_{+})]^2\sigma(|\varphi_+|)+\frac{1}{2}\right]
		\right.  \\
		&&\nonumber   +\Gamma^{-}\left[[\mathbf{Re}(\varphi_{-})]^2\beta(|\varphi_-|)+[\mathbf{Im}(\varphi_{-})]^2\sigma(|\varphi_-|)+\frac{1}{2}\right]
		\\
		&&\nonumber  +2\mathbf{Re}\left(\frac{\Gamma^{+-}}{2\sqrt{_0F_2(1,1;|\varphi_{+}|^2)}\sqrt{_0F_2(1,1;|\varphi_{-}|^2)}}\right.\\
		&&\nonumber   \times\left.\left.\left[\frac{(\varphi_{-}^2+\varphi^{\ast2}_+)}{2}\;_0F_2(1,3;\varphi^{\ast}_{+}\varphi_{-})+2\varphi^{\ast}_+\varphi_-\;_0F_2(2,2;\varphi^{\ast}_{+}\varphi_{-})+\;_0F_2(1,1;\varphi^{\ast}_{+}\varphi_{-})\right]
		\right)\right\} \\
		&&\nonumber   -\Delta_1^{-2}\left\{\sqrt{2}\Gamma^{+}\mathbf{Re}(\varphi_{+})\frac{_0F_2(1,2;|\varphi_+|^2)}{_0F_2(1,1;|\varphi_+|^2)}+\sqrt{2}\Gamma^{-}\mathbf{Re}(\varphi_{-})\frac{_0F_2(1,2;|\varphi_-|^2)}{_0F_2(1,1;|\varphi_-|^2)}\right. \\
		&& \left.+\sqrt{2}\mathbf{Re}\left((\varphi_{-}+\varphi^{\ast}_+)\frac{\Gamma^{+-}\;_0F_2(1,2;\varphi^{\ast}_{+}\varphi_{-})}{\sqrt{_0F_2(1,1;|\varphi_{+}|^2)}\sqrt{_0F_2(1,1;|\varphi_{-}|^2)}}\right)\right\}^2,\\
		\nonumber  \sigma^{2 (\mathrm{nl})}_p&=& \Delta_1^{-1}\left\{\Gamma^{+}\left[[\mathbf{Re}(\varphi_{+})]^2\sigma(|\varphi_+|)+[\mathbf{Im}(\varphi_{+})]^2\beta(|\varphi_+|)+\frac{1}{2}\right]
		\right.  \\
		&&\nonumber   +\Gamma^{-}\left[[\mathbf{Re}(\varphi_{-})]^2\sigma(|\varphi_-|)+[\mathbf{Im}(\varphi_{-})]^2\beta(|\varphi_-|)+\frac{1}{2}\right]\\
		&&\nonumber   +2\mathbf{Re}\left(\frac{\Gamma^{+-}}{2\sqrt{_0F_2(1,1;|\varphi_{+}|^2)}\sqrt{_0F_2(1,1;|\varphi_{-}|^2)}}\right.\\
		&&\nonumber   \times\left.\left.\left[-\frac{(\varphi_{-}^2+\varphi^{\ast2}_{+})}{2}\;_0F_2(1,3;\varphi^{\ast}_{+}\varphi_{-})+2\varphi^{\ast}_+\varphi_-\;_0F_2(2,2;\varphi^{\ast}_{+}\varphi_{-})+\;_0F_2(1,1;\varphi^{\ast}_{+}\varphi_{-})\right]
		\right)\right\} \\
		&&\nonumber   -\Delta_1^{-2}\left\{\sqrt{2}\Gamma^{+}\mathbf{Im}(\varphi_{+})\;\frac{_0F_2(1,2;|\varphi_+|^2)}{_0F_2(1,1;|\varphi_+|^2)}+\sqrt{2}\Gamma^{-}\mathbf{Im}(\varphi_{-})\;\frac{_0F_2(1,2;|\varphi_-|^2)}{_0F_2(1,1;|\varphi_-|^2)}\right. \\
		&& \left.+\sqrt{2}\mathbf{Re}\left(-i(\varphi_{-}-\varphi^{\ast}_+)\frac{\Gamma^{+-}\;_0F_2(1,2;\varphi^{\ast}_{+}\varphi_{-})}{\sqrt{_0F_2(1,1;|\varphi_{+}|^2)}\sqrt{_0F_2(1,1;|\varphi_{-}|^2)}}\right)\right\}^2,
	\end{eqnarray}
	\normalsize
	where $\beta(|\varphi_{\pm}|)$ and $\sigma(|\varphi_{\pm}|)$ are given by eq. (\ref{24.3}), and
	\begin{eqnarray*}
		\Gamma^{+} &=&|\gamma_{1+}|^2+|\gamma_{2+}Y|^2, \\
		\Gamma^{-}&=&|\gamma_{1-}|^2+|\gamma_{2-}Y|^2, \\ \Gamma^{+-}&=&\gamma^{\ast}_{1+}\gamma_{1-}+\gamma^{\ast}_{2+}\gamma_{2-}|Y|^2, \\
		\Delta_1 &=& \Gamma^{+}
		+\Gamma^{-}
		+2\mathbf{Re}\left(\frac{\Gamma^{+-}\;_0F_2(1,1;\varphi^{\ast}_{+}\varphi_-)}{\sqrt{_0F_2(1,1;|\varphi_{+}|^2)}\sqrt{_0F_2(1,1;|\varphi_-|^2)}}\right).
	\end{eqnarray*}
	
	\subsection{\label{subsec:level42}Second deformation of the SAO}
	Let us make the change $\hat{a}\rightarrow\hat{N}\hat{a}$ in eq. (\ref{39}) so that the deformed annihilation operator $\hat{A}''$ becomes now:
	\begin{equation}\label{53}
		\hat{A}''=\left(
		\begin{array}{cc}
			k_1\hat{N}\hat{a} & k_2 \\
			k_3(\hat{N}\hat{a})^2 & k_4\hat{N}\hat{a} \\
		\end{array}
		\right).
	\end{equation}
	
	Let $|\mathcal{Z}\rangle$ be the eigenstate of the operator $\hat{A}''$, such that
	\begin{equation}\label{54}
		\hat{A}''|\mathcal{Z}\rangle=Z|\mathcal{Z}\rangle, \quad |\mathcal{Z}\rangle=
		\sum\limits_{n=0}^{\infty} u_n \vert\Psi_n^+\rangle + \sum\limits_{n=1}^{\infty} v_n \vert\Psi_n^-\rangle .
	\end{equation}
	Using this eigenvalue equation, the following recurrence relations are obtained
	\begin{equation}\label{55}
		\left\{
		\begin{array}{c}
			k_1n\sqrt{n+1}u_{n+1}+k_2v_{n+1}=Zu_{n}, \\
			k_3n(n-1)\sqrt{n}\sqrt{n+1}u_{n+1}+k_4(n-1)\sqrt{n}v_{n+1}=Zv_n. \\
		\end{array}
		\right.
	\end{equation}
	
	From the expressions in eq. (\ref{55}), we obtain first that
	\begin{eqnarray}\label{56}
		\nonumber  n=0 &\longrightarrow& Zu_0=k_2v_1,
		\\
		n=1 &\longrightarrow& v_1=0, \quad u_0=0, \quad \sqrt{2}k_1u_2=Zu_1-k_2v_2,
	\end{eqnarray}
	whereby $u_1$ and $v_2$ are the free parameters.
	
	If we define now $\tilde{u}_{n}=(n-1)!\sqrt{n!}Z^{1-n}u_n$ and $\tilde{v}_n=(n-2)!\sqrt{(n-1)!}Z^{1-n}v_n$ for $n\geq2$, then the system in eq. (\ref{55}) can be expressed as
	\begin{equation}\label{57}
		K\left(
		\begin{array}{c}
			\tilde{u}_{n+1} \\
			\tilde{v}_{n+1} \\
		\end{array}
		\right)=\left(
		\begin{array}{c}
			\tilde{u}_{n} \\
			\tilde{v}_{n} \\
		\end{array}
		\right),
	\end{equation}
	where $K$ is once again the matrix defined by eq. (\ref{40.41}). For the nonzero eigenvalues  $\kappa_{\pm}$ of the matrix $K$, we define the quantities $\phi_{\pm}\equiv Z\kappa_{\pm}^{-1}$.
	
	As in the previous case, the same classification of the nonlinear supercoherent states, according to the values taken by $\kappa_{\pm}$, is obtained.
	
	\subsubsection{Classification of the deformed families}
	\paragraph{\textbf{Generic}.} As before, the solution for the matrix equation (\ref{57}) turns out to be
	\begin{equation}\label{59}
		\left(
		\begin{array}{c}
			u_{n+1} \\
			v_{n+1} \\
		\end{array}
		\right)=u_1\left(
		\begin{array}{c}
			\frac{Z}{k_1n!\sqrt{(n+1)!}}d_{11} \\
			\frac{Z}{k_1(n-1)!\sqrt{n!}}d_{21} \\
		\end{array}
		\right)+v_2\left(
		\begin{array}{c}
			\frac{1}{n!\sqrt{(n+1)!}}\left(d_{12}-d_{11}\frac{k_2}{k_1}\right) \\
			\frac{1}{(n-1)!\sqrt{n!}}\left(d_{22}-d_{21}\frac{k_2}{k_1}\right) \\
		\end{array}
		\right),
	\end{equation}
	where
	\footnotesize
	\begin{equation*}\label{60}
		\left(
		\begin{array}{cc}
			d_{11} & d_{12} \\
			d_{21} & d_{22} \\
		\end{array}
		\right)=\frac{1}{\kappa_{+}-\kappa_{-}}\left(
		\begin{array}{cc}
			(\kappa_{+}-k_4)\phi_{+}^{n-1}-(\kappa_{-}-k_4)\phi_{-}^{n-1} & k_2(\phi_{+}^{n-1}-\phi_{-}^{n-1}) \\
			k_3(\phi_{+}^{n-1}-\phi_{-}^{n-1}) & (\kappa_{+}-k_4)\phi_{-}^{n-1}-(\kappa_{-}-k_4)\phi_{+}^{n-1} \\
		\end{array}
		\right).
	\end{equation*}
	 \normalsize
	
	The supercoherent states $|\mathcal{Z}\rangle$ become now
	\begin{equation}\label{61}
		|\mathcal{Z}\rangle=u'_1|\mathcal{Z}_\mathrm{A}\rangle+v'_2|\mathcal{Z}_\mathrm{C}\rangle,
	\end{equation}
	where $u'_1=u_1k_1^{-1}$, $v'_2=v_2(Zk_1)^{-1}$ and
	\begin{eqnarray}\label{62}
		|\mathcal{Z}_\mathrm{A}\rangle&=&\frac{1}{\kappa_{+}-\kappa_{-}}G_\mathrm{A}^{\mathrm{NL}}\left(
		\begin{array}{c}
			|\phi_{+}\rangle_{\mathrm{NL}} \\
			|\phi_{-}\rangle_{\mathrm{NL}} \\
		\end{array}
		\right), \\ |\mathcal{Z}_\mathrm{C}\rangle&=&\frac{1}{\kappa_{+}-\kappa_{-}}G_\mathrm{C}^{\mathrm{NL}}\left(
		\begin{array}{c}
			|\phi_{+}\rangle_{\mathrm{NL}} \\
			|\phi_{-}\rangle_{\mathrm{NL}} \\
		\end{array}
		\right),
	\end{eqnarray}
	with
	\begin{eqnarray*}\label{63}
		G_\mathrm{A}^{\mathrm{NL}} &=& 
		\left(
		\begin{array}{cc}
			\kappa_{+}(\kappa_{+}-k_4) & -\kappa_{-}(\kappa_{-}-k_4) \\
			k_3Z & -k_3Z \\
		\end{array}
		\right),
		\\
		G_\mathrm{C}^{\mathrm{NL}} &=&  
		\left(
		\begin{array}{cc}
			k_2\kappa_{+}\kappa_{-} & -k_2\kappa_{+}\kappa_{-} \\
			Z[k_1\kappa_{+}-(k_1^2+k_2k_3)] & -Z[k_1\kappa_{-}-(k_1^2+k_2k_3)] \\
		\end{array}
		\right).
	\end{eqnarray*}
	
	Again, from the set $\{|\mathcal{Z}_\mathrm{A}\rangle,|\mathcal{Z}_\mathrm{C}\rangle\}$ we can build a new one formed by the elements
	\begin{equation}\label{63.1}
		|\mathcal{Z}_{\pm}\rangle_{\mathrm{NL}}=\left(
		\begin{array}{c}
			k_2\kappa_{\pm}|\phi_{\pm}\rangle_{\mathrm{NL}} \\
			(\kappa_{\pm}-k_1)Z|\phi_{\pm}\rangle_{\mathrm{NL}} \\
		\end{array}
		\right),
	\end{equation}
	\textit{i.e.}, we pass from the parameter space $\{k_1,k_2,k_3,k_4\}$ to the new one $\{\kappa_{+},\kappa_{-},k_1,k_2\}$.
	
	\paragraph{\textbf{Degenerate}.} The deformed supercoherent states are explicitly given by
	\begin{equation}\label{64}
		|\mathcal{Z}\rangle=u'_1|\mathcal{Z}^\mathrm{d}_\mathrm{A}\rangle_{\mathrm{NL}}+v'_2|\mathcal{Z}^\mathrm{d}_\mathrm{C}\rangle_{\mathrm{NL}},
	\end{equation}
	where
	\begin{eqnarray}\label{65}
		|\mathcal{Z}^\mathrm{d}_\mathrm{A}\rangle_{\mathrm{NL}}&=&G_\mathrm{A}^{d,\mathrm{NL}}\left(
		\begin{array}{c}
			|\phi\rangle_{\mathrm{NL}} \\
			|\phi'\rangle_{\mathrm{NL}} \\
		\end{array}
		\right), \\
		|\mathcal{Z}^\mathrm{d}_\mathrm{C}\rangle_{\mathrm{NL}}&=&G_\mathrm{C}^{\mathrm{d},\mathrm{NL}}\left(
		\begin{array}{c}
			|\phi\rangle_{\mathrm{NL}} \\
			|\phi'\rangle_{\mathrm{NL}} \\
		\end{array}
		\right),
	\end{eqnarray}
	with $|\phi'\rangle=\frac{d}{d\phi}|\phi\rangle$ and
	\begin{equation*}\label{66}
		G_\mathrm{A}^{\mathrm{d},\mathrm{NL}} = 
		\left(
		\begin{array}{cc}
			k_1 & -(\kappa-k_4)\phi \\
			0 & -k_3\phi^2 \\
		\end{array}
		\right),
		\quad
		G_\mathrm{C}^{\mathrm{d},\mathrm{NL}} =  
		\kappa\left(
		\begin{array}{cc}
			0 & -k_2\phi \\
			k_1\phi & -(\frac{k_4-k_1}{2})\phi^2 \\
		\end{array}
		\right).
	\end{equation*}
	
	\paragraph{\textbf{Singular}.} The corresponding deformed supercoherent state is now
	\begin{equation}\label{67}
		|\mathcal{Z}\rangle_{\mathrm{NL}}=\left(
		\begin{array}{c}
			k_1|\phi\rangle_{\mathrm{NL}} \\
			k_3\phi|\phi\rangle_{\mathrm{NL}} \\
		\end{array}
		\right),
	\end{equation}
	with $\phi\equiv Z(k_1+k_4)^{-1}$.
	
	\subsubsection{Superposition and deformed uncertainties.}
	Consider again the following superposition of states $|\mathcal{Z}_{\pm}\rangle_{\mathrm{NL}}$ of eq.~(\ref{63.1}) with parameters $\eta$ and $\lambda$:
	\begin{eqnarray}\label{68}
		|\mathcal{Z}_m\rangle_{\mathrm{NL}} &=& \cos\eta|\mathcal{Z}_+\rangle_{\mathrm{NL}}+\exp(i\lambda)\sin\eta|\mathcal{Z}_-\rangle_{\mathrm{NL}} \nonumber\\
		&=& \left(
		\begin{array}{c}
			\gamma_{1+}|\phi_+\rangle_{\mathrm{NL}}+\gamma_{1-}|\phi_-\rangle_{\mathrm{NL}} \\
			\gamma_{2+}|\phi_+\rangle_{\mathrm{NL}}+\gamma_{2-}|\phi_-\rangle_{\mathrm{NL}} \\
		\end{array}
		\right),
	\end{eqnarray}
	where the quantities $\gamma_{1\pm}$ and $\gamma_{2\pm}$ are defined before, in eqs. (\ref{45a}) and (\ref{45b}).
	
	In order to find the uncertainties of the position and momentum operators in the nonlinear supercoherent state of eq.~(\ref{68}), we will use now the following expressions involving $|\alpha_1\rangle_{\mathrm{NL}}$ and $|\alpha_2\rangle_{\mathrm{NL}}$:
	\begin{eqnarray}\label{69}
		\langle\alpha_1|\alpha_2\rangle_{\mathrm{NL}} &=& \frac{_0F_2(1,2;\alpha^{\ast}_1\alpha_2)}{\sqrt{_0F_2(1,2;|\alpha_1|^2)}\sqrt{_0F_2(1,2;|\alpha_2|^2)}}
		,\\
		\langle\alpha_1|\hat{x}|\alpha_2\rangle_{\mathrm{NL}} &=& \frac{\alpha_2+\alpha^{\ast}_1}{\sqrt{2}}\frac{_0F_2(2,2;\alpha^{\ast}_1\alpha_2)}{\sqrt{_0F_2(1,2;|\alpha_1|^2)}\sqrt{_0F_2(1,2;|\alpha_2|^2)}}, \\
		\nonumber  \langle\alpha_1|\hat{x}^2|\alpha_2\rangle_{\mathrm{NL}} &=& \frac{1}{2\sqrt{_0F_2(1,2;|\alpha_1|^2)}\sqrt{_0F_2(1,2;|\alpha_2|^2)}} \\
		&& \times \left[\frac{(\alpha_2+\alpha^{\ast}_1)^2}{2}\,_0F_2(2,3;\alpha^{\ast}_1\alpha_2)+3\,_0F_2(1,2;\alpha^{\ast}_1\alpha_2)\right], \\
		\langle\alpha_1|\hat{p}\:|\alpha_2\rangle_{\mathrm{NL}} &=& \frac{\alpha_2-\alpha^{\ast}_1}{\sqrt{2}\:i}\frac{_0F_2(2,2;\alpha^{\ast}_1\alpha_2)}{\sqrt{_0F_2(1,2;|\alpha_1|^2)}\sqrt{_0F_2(1,2;|\alpha_2|^2)}}, \\
		\nonumber  \langle\alpha_1|\hat{p}^2|\alpha_2\rangle_{\mathrm{NL}} &=& \frac{1}{2\sqrt{_0F_2(1,2;|\alpha_1|^2)}\sqrt{_0F_2(1,2;|\alpha_2|^2)}} \\
		&& \times \left[-\frac{(\alpha_2-\alpha^{\ast}_1)^2}{2}\,_0F_2(2,3;\alpha^{\ast}_1\alpha_2)+3\,_0F_2(1,2;\alpha^{\ast}_1\alpha_2)\right].
	\end{eqnarray}
	When $|\alpha_1\rangle_{\mathrm{NL}}=|\alpha_2\rangle_{\mathrm{NL}}\equiv|\alpha\rangle_{\mathrm{NL}}$ they reduce now to those of eqs. (\ref{17a})-(\ref{17d}).
	
	The expressions for the uncertainties are finally:
	\begin{eqnarray}\label{70}
		\sigma^{2 (\mathrm{NL})}_x &=& \Delta_2^{-1}\left\{\Gamma^{+}\left[[\mathbf{Re}(\phi_{+})]^2\frac{_0F_2(2,3;|\phi_+|^2)}{_0F_2(1,2;|\phi_+|^2)}+\frac{3}{2}\right]
		\right.  \nonumber\\
		&& +\Gamma^{-}\left[[\mathbf{Re}(\phi_{-})]^2\frac{_0F_2(2,3;|\phi_-|^2)}{_0F_2(1,2;|\phi_-|^2)}+\frac{3}{2}\right]\nonumber \\
		&&  +2\mathbf{Re}\left(\frac{\Gamma^{+-}}{2\sqrt{_0F_2(1,2;|\phi_{+}|^2)}\sqrt{_0F_2(1,2;|\phi_{-}|^2)}}\right.\nonumber\\
		&&  \times\left.\left.\left[\frac{(\phi_{-}+\phi^{\ast}_+)^2}{2}\;_0F_2(2,3;\phi^{\ast}_{+}\phi_{-})+3\,_0F_2(1,2;\phi^{\ast}_{+}\phi_{-})\right]
		\right)\right\} \nonumber \\
		&& -\Delta_2^{-2}\left\{\sqrt{2}\,\Gamma^{+}\mathbf{Re}(\phi_{+})\frac{_0F_2(2,2;|\phi_+|^2)}{_0F_2(1,2;|\phi_+|^2)}+\sqrt{2}\,\Gamma^{-}\mathbf{Re}(\phi_{-})\frac{_0F_2(2,2;|\phi_-|^2)}{_0F_2(1,2;|\phi_-|^2)}\right. \nonumber \\
		&& \left.+\sqrt{2}\,\mathbf{Re}\left((\phi_{-}+\phi^{\ast}_+)\frac{\Gamma^{+-}\;_0F_2(2,2;\phi^{\ast}_{+}\phi_{-})}{\sqrt{_0F_2(1,2;|\phi_{+}|^2)}\sqrt{_0F_2(1,2;|\phi_{-}|^2)}}\right)\right\}^2,\\
		\sigma^{2 (\mathrm{NL})}_p&=& \Delta_2^{-1}\left\{\Gamma^{+}\left[[\mathbf{Im}(\phi_{+})]^2\frac{_0F_2(2,3;|\phi_+|^2)}{_0F_2(1,2;|\phi_+|^2)}+\frac{3}{2}\right]
		\right. \nonumber \\
		&& +\Gamma^{-}\left[[\mathbf{Im}(\phi_{-})]^2\frac{_0F_2(2,3;|\phi_-|^2)}{_0F_2(1,2;|\phi_-|^2)}+\frac{3}{2}\right]\nonumber\\
		&& +2\mathbf{Re}\left(\frac{\Gamma^{+-}}{2\sqrt{_0F_2(1,2;|\phi_{+}|^2)}\sqrt{_0F_2(1,2;|\phi_{-}|^2)}}\right. \nonumber\\
		&& \times\left.\left.\left[-\frac{(\phi_{-}-\phi^{\ast}_+)^2}{2}\;_0F_2(2,3;\phi^{\ast}_{+}\phi_{-})+3\,_0F_2(1,2;\phi^{\ast}_{+}\phi_{-})\right]
		\right)\right\} \nonumber \\
		&& -\Delta_2^{-2}\left\{\sqrt{2}\,\Gamma^{+}\mathbf{Im}(\phi_{+})\frac{_0F_2(2,2;|\phi_+|^2)}{_0F_2(1,2;|\phi_+|^2)}+\sqrt{2}\,\Gamma^{-}\mathbf{Im}(\phi_{-})\frac{_0F_2(2,2;|\phi_-|^2)}{_0F_2(1,2;|\phi_-|^2)}\right. \nonumber\\
		&& \left.+\sqrt{2}\,\mathbf{Re}\left(-i(\phi_{-}-\phi^{\ast}_+)\frac{\Gamma^{+-}\;_0F_2(2,2;\phi^{\ast}_{+}\phi_{-})}{\sqrt{_0F_2(1,2;|\phi_{+}|^2)}\sqrt{_0F_2(1,2;|\phi_{-}|^2)}}\right)\right\}^2
	\end{eqnarray}
	where now
	\begin{eqnarray*}
		\Gamma^{+}&=&|\gamma_{1+}|^2+|\gamma_{2+}Z|^2, \\
		\Gamma^{-}&=&|\gamma_{1-}|^2+|\gamma_{2-}Z|^2, \\ \Gamma^{+-}&=&\gamma^{\ast}_{1+}\gamma_{1-}+\gamma^{\ast}_{2+}\gamma_{2-}|Z|^2, \\
		\Delta_2 &=& \Gamma^{+}
		+\Gamma^{-}
		+2\mathbf{Re}\left(\frac{\Gamma^{+-}\;_0F_2(1,2;\phi^{\ast}_{+}\phi_-)}{\sqrt{_0F_2(1,2;|\phi_{+}|^2)}\sqrt{_0F_2(1,2;|\phi_-|^2)}}\right).
	\end{eqnarray*}
	
	\subsection{\label{subsec:level43}A particular case}
	
	\begin{figure*}[t]
		\centering
		\includegraphics[width=12pc]{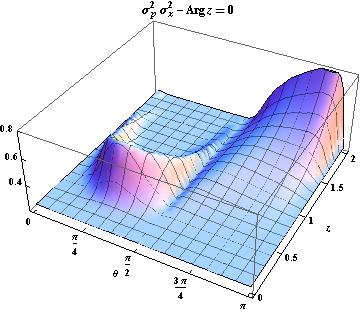} \hspace{2pc}
		\includegraphics[width=12pc]{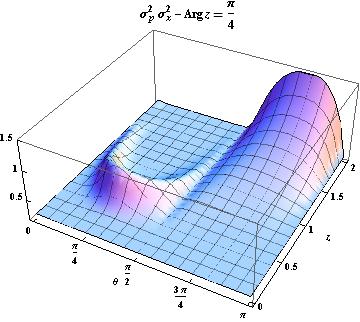}
		\caption{\label{fig:nl3}Uncertainty relation $\sigma^2_x\sigma^2_p$ corresponding to the supercoherent states derived in \cite{zypman}, in the interval $0<\theta<\pi$.}
	\end{figure*}
	Let the annihilation operators $\hat{A}_{\theta}'$ and $\hat{A}_{\theta}''$ be given respectively by
	\begin{eqnarray}\label{73}
		\hat{A}_{\theta}'&=&\left(
		\begin{array}{cc}
			(\hat{N}+\hat{1})\hat{a} & \cos\theta \\
			\tiny{ }[(\hat{N}+\hat{1})\hat{a}]^2\sin\theta & (\hat{N}+\hat{1})\hat{a} \\
		\end{array}
		\right), \\
		\hat{A}_{\theta}''&=&\left(
		\begin{array}{cc}
			\hat{N}\hat{a} & \cos\theta \\
			(\hat{N}\hat{a})^2\sin\theta & \hat{N}\hat{a} \\
		\end{array}
		\right),
	\end{eqnarray}
	for $0\leq\theta\leq\pi$, which correspond to two deformations of the SAO of \cite{zypman}. The eigenvalues $\kappa_{\pm}$ of the matrix $K$ of eq. (\ref{40.41}) become now:
	\begin{equation}\label{73.0}
		\kappa_{\pm}=1\pm\sqrt{\frac{1}{2}\sin(2\theta)},
	\end{equation}
	so that for $0<\theta<\pi/2$ both $\kappa_{\pm}$ are real while for $\pi/2<\theta<\pi$ they are complex.
	
	Consider first the nonlinear supercoherent states of 
	eq. (\ref{52.5}) with $\eta=\lambda=\pi/4$ (\textit{i.e.}, those associated to $\hat{A}'_{\theta}$):
	\begin{equation}\label{73.1}
		|\mathcal{Y}_{\theta}\rangle_{\mathrm{nl}}=\frac{1}{\sqrt{2}}(|\mathcal{Y}_{+,\theta}\rangle_{\mathrm{nl}}+\exp(i\pi/4)|\mathcal{Y}_{-,\theta}\rangle_{\mathrm{nl}}).
	\end{equation}
	
	In addition, we will analyze also the nonlinear supercoherent states of eq. (\ref{68}) with the same $\eta$ and $\lambda$ (the ones associated to $\hat{A}_{\theta}''$):
	\begin{equation}\label{74}
		|\mathcal{Z}_{\theta}\rangle_{\mathrm{NL}}=\frac{1}{\sqrt{2}}(|\mathcal{Z}_{+,\theta}\rangle_{\mathrm{NL}}+\exp(i\pi/4)|\mathcal{Z}_{-,\theta}\rangle_{\mathrm{NL}}).
	\end{equation}
	
	\begin{figure*}
		\centering
		\includegraphics[width=12pc]{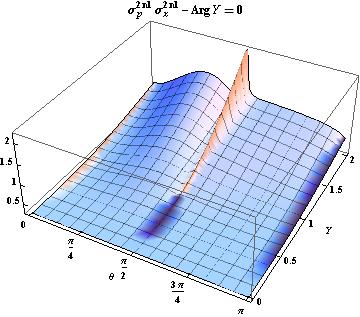} \hspace{2pc}
		\includegraphics[width=12pc]{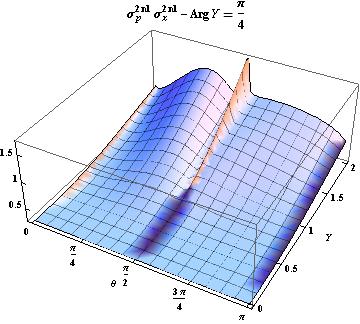}
		\caption{\label{fig:nl4}Uncertainty relation $\sigma^{2 (\mathrm{nl})}_{x}\sigma^{2 (\mathrm{nl})}_{p}$ for the supercoherent states $|\mathcal{Y}_{\theta}\rangle_{\mathrm{nl}}$, in the interval $0<\theta<\pi$. For real $Y$ (Arg$[Y]=0$) it is shown on the left side while the right side illustrates it for complex $Y$, with Arg$[Y]=\pi/4$.}
	\end{figure*}
	
\begin{figure*}
			\centering
			\includegraphics[width=12pc]{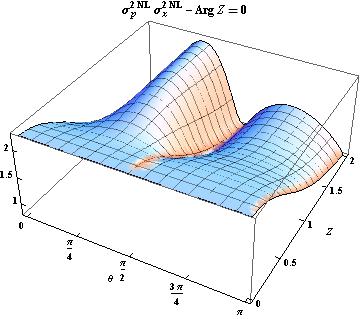} \hspace{2pc}
			\includegraphics[width=12pc]{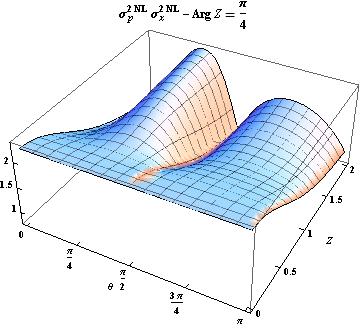}
			\caption{\label{fig:nl5}Uncertainty relation $\sigma^{2 (\mathrm{NL})}_{x}\sigma^{2 (\mathrm{NL})}_{p}$ for the supercoherent states $|\mathcal{Z}_{\theta}\rangle_{\mathrm{NL}}$, in the interval $0<\theta<\pi$. For real $Z$ (Arg$[Z]=0$) it is shown on the left side while the right side illustrates it for complex $Z$, with Arg$[Z]=\pi/4$.}
		\end{figure*}

	For comparison, the square of the Heisenberg uncertainty relation associated with the linear supercoherentes states of \cite{zypman} is shown in fig. \ref{fig:nl3}. For $z$ real (left plot) this quantity reaches a maximum value equal to $0.83$ at $|z|\sim0.5$ for $\kappa_{\pm}$ real ($0<\theta<\pi/2$), while it shows a growing behavior as $|z|$ increases for $\kappa_{\pm}$ complex ($\pi/2<\theta<\pi$). Moreover, for complex $z$ (right plot), the square of the uncertainty relation behaves similarly as for $z$ real.
	Meanwhile, figs. \ref{fig:nl4} and \ref{fig:nl5} show the same quantities $\sigma_{x}^{2(\mathrm{nl})}\sigma_{p}^{2(\mathrm{nl})}$ and $\sigma_{x}^{2(\mathrm{NL})}\sigma_{p}^{2(\mathrm{NL})}$ associated to the nonlinear supercoherent states of eqs.~(\ref{73.1}) and (\ref{74}), respectively. By comparing the results of figs. \ref{fig:nl3}, \ref{fig:nl4} and \ref{fig:nl5}, it is clear that the uncertainty relation behaves differently, depending if either it is calculated for the linear supercoherentes states $|\mathcal{Z}_{\theta}\rangle$ of \cite{zypman} or for the deformed supercoherent ones of eqs. (\ref{73.1}) and (\ref{74}).
	
	Figure \ref{fig:nl4} indicates that for the nonlinear coherent states $|\mathcal{Y}_{\theta}\rangle_{\mathrm{nl}}$ of eq.~(\ref{73.1}) the uncertainty relation reaches its maximum at the vicinity of $\theta=\pi/2$, which marks the boundary between the real and complex eigenvalues $\kappa_\pm$ (degenerate case). On the other hand, fig. \ref{fig:nl5} shows the opposite behavior, namely, the uncertainty relation decreases around the same $\theta$ value. Also, the uncertainty square for the nonlinear coherent states $|\mathcal{Y}_{\theta}\rangle_{\mathrm{nl}}$ starts from the minimum value $1/4$ for $Y=0$ and then increases slowly as $|Y|$ grows, while for the states $|\mathcal{Z}_{\theta}\rangle_{\mathrm{NL}}$ it departs from the value $9/4$ at $Z=0$, then it decreases for certain interval of $|Z|$ in order to increase finally for sufficiently large $|Z|$.
	
	Let us note that, although eqs.~(\ref{73.1}) and (\ref{74}) at first sight look similar as the expressions for the even and odd coherent states (see e.g. \cite{dmm74,mm01}), these superpositions cannot be considered as the appropriate generalization of such states. In fact, the coherent states $\vert\mathcal{Y}_{\pm,\theta}\rangle_{\mathrm{nl}}$ ($\vert\mathcal{Z}_{\pm,\theta}\rangle_{\mathrm{NL}}$) are eigenvectors of the SAO $\hat{A}'_\theta$ ($\hat{A}''_\theta$) with the same eigenvalue $Y$ ($Z$) while the even and odd coherent states are superpositions of coherent states with different eigenvalues. In order to determine the even and odd coherent states for the SUSY harmonic oscillator we must look for the eigenstates of $\hat{A}'^2_\theta$ and $\hat{A}''^2_\theta$, which is a work currently in progress.
	
	\section{\label{sec:level5}Geometric phases}
	As it was shown in section \ref{sec:level3}, the SUSY harmonic oscillator executes an evolution loop, thus
	any state turns out to be cyclic and then it has associated a geometric phase. In particular, this is valid for the
	nonlinear supercoherent states. Let us make next the explicit calculation of their corresponding
	geometric phases.
	
	\subsection{\label{subsec:level51}First deformation of the SAO}
\begin{figure*}
		\centering
		\includegraphics[width=12pc]{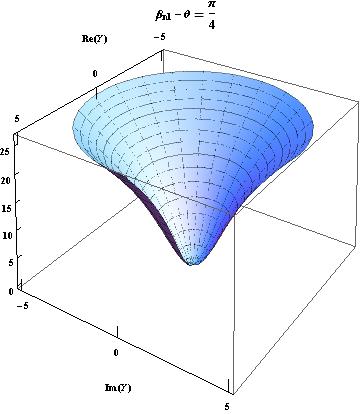} \hspace{2pc}
		\includegraphics[width=12pc]{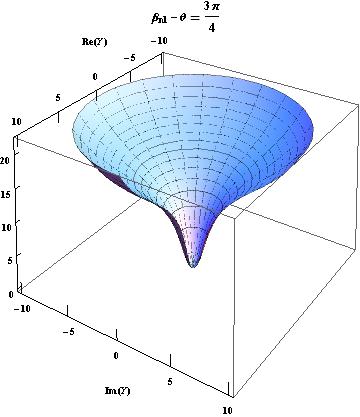}
		\caption{\label{fig:nl6} Geometric phase $\beta_{\mathrm{nl}}$ for the supercoherent states $|\mathcal{Y}_{\theta}\rangle_{\mathrm{nl}}$ for $\theta=\pi/4$ (left) and $\theta=3\pi/4$ (right).}
	\end{figure*}

	The states $|\mathcal{Y}_\pm\rangle_{\mathrm{nl}}$ are cyclic, with $\tau = 2\pi/\omega$ and $\Omega = 0$. Thus, using eq. (\ref{gptih}) it
	is obtained the following expression for $\beta_{\mathrm{nl}}$ (see fig. \ref{fig:nl6}):
	\small
	\begin{eqnarray}\label{81}
		\nonumber	\beta_{\mathrm{nl}}=\Omega+\tau\langle\hat{H}\rangle_{\mathrm{nl}}&=&2\pi\Delta_1^{-1}\bigg\{\Gamma^+\frac{_0F_2(2,2;| \varphi_+|^2)}{_0F_2(1,1;|\varphi_+|^2)}|\varphi_+|^2+\Gamma^-\frac{_0F_2(2,2;| \varphi_-|^2)}{_0F_2(1,1;|\varphi_-|^2)}|\varphi_-|^2\\
\nonumber		&&\quad	+2\mathbf{Re}\left(\Gamma^{+-}\frac{_0F_2(2,2;\varphi_+^\ast\varphi_-)}{\sqrt{_0F_2(1,1;|\varphi_+|^2)\,_0F_2(1,1;|\varphi_-|^2)}}\varphi_+^\ast\varphi_-\right)\\
		&&\quad+|\gamma_{2+}Y|^2+|\gamma_{2-}Y|^2+2|Y|^2\mathbf{Re}(\gamma_{2+}^\ast\gamma_{2-})\bigg\}.
	\end{eqnarray}
	\normalsize
	
	\subsection{\label{subsec:level52}Second deformation of the SAO}
	
\begin{figure*}
		\centering
		\includegraphics[width=12pc]{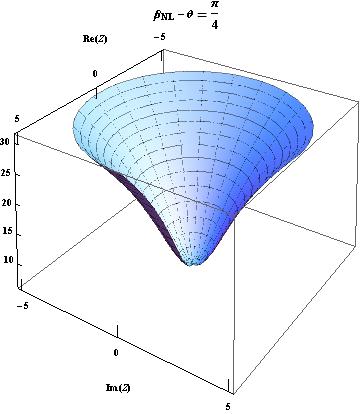} \hspace{2pc}
		\includegraphics[width=12pc]{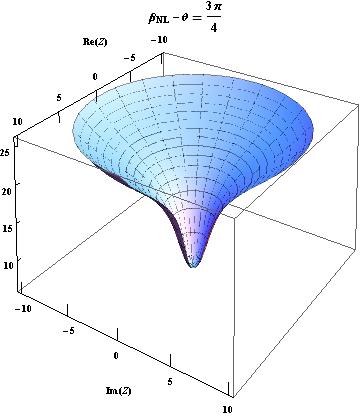}
		\caption{\label{fig:nl7}Geometric phase $\beta_{\mathrm{NL}}$ for the supercoherent states $|\mathcal{Z}_{\theta}\rangle_{\mathrm{NL}}$ for $\theta=\pi/4$ (left) and $\theta=3\pi/4$ (right).}
	\end{figure*}

	The states $|\mathcal{Z}_\pm\rangle_{\mathrm{NL}}$ are also cyclic with $\tau = 2\pi/\omega$ and $\Omega = 0$. Using eq. (\ref{gptih})
	it is obtained now the following geometric phase $\beta_{\mathrm{NL}}$ (see fig. \ref{fig:nl7}):
	
	\small
	\begin{eqnarray}\label{86}
		\nonumber	\beta_{\mathrm{NL}}=\Omega+\tau\langle\hat{H}\rangle_{\mathrm{NL}}&=&2\pi\Delta_2^{-1}\bigg\{\Gamma^+\left[\frac{|\phi_+|^2}{2}\frac{_0F_2(2,3;| \phi_+|^2)}{_0F_2(1,2;|\phi_+|^2)}+1\right]\\
		\nonumber&&\quad+\Gamma^-\left[\frac{|\phi_-|^2}{2}\frac{_0F_2(2,3;| \phi_-|^2)}{_0F_2(1,2;|\phi_-|^2)}+1\right]\\
		\nonumber&&\quad	+2\mathbf{Re}\left(\frac{\Gamma^{+-}}{\sqrt{_0F_2(1,2;|\phi_+|^2)\,_0F_2(1,2;|\phi_-|^2)}}\right.\\
		\nonumber &&\quad\left.\times\left[\frac{\phi_+^\ast\phi_-}{2}\,_0F_2(2,3;\phi_+^\ast\phi_-)+\,_0F_2(1,2;\phi_+^\ast\phi_-)\right]\right)\\
	&&\quad+|\gamma_{2+}Z|^2+|\gamma_{2-}Z|^2+2|Z|^2\mathbf{Re}(\gamma_{2+}^\ast\gamma_{2-})\bigg\}
	\end{eqnarray}
	\normalsize
	
	Figures \ref{fig:nl6} and \ref{fig:nl7} show that the behavior of the geometric phases $\beta_{\mathrm{nl}}$ and $\beta_{\mathrm{NL}}$ depends strongly on the parameter $\theta$. In fact, both geometric phases grow quickly for $\theta=\pi/4$ while for $\theta=3\pi/4$ they increase more slowly but with a less smooth character. On the other hand, in the limit $Y\rightarrow0$ and $Z\rightarrow0$, both $\beta_{\mathrm{nl}}$ and $\beta_{\mathrm{NL}}$ reach the minimum value $0$ and $2\pi$, respectively, which is consistent with the fact that the mean value of any Hamiltonian is at least equal to the minimum energy eigenvalue, \textit{i.e.}, $\langle\hat{H}\rangle\geq E_\text{min}$, where $E_\text{min}$ denotes the smallest energy involved in the NLSCS.
	
	\section{\label{sec:level6}Concluding remarks}
	If instead of $\hat{a}$ we set in the harmonic oscillator algebra as annihilation operator either $\tilde{a}=(\hat{N}+\hat{1})\hat{a}$ or $\tilde{a}=\hat{N}\hat{a}$, important consequences arise concerning the form of their respective eigenvectors (coherent states) as well as the associated algebraic structure. While for the usual operator $\hat{a}$ its eigenstates have the form given in eq.~(\ref{13}), the ones corresponding to $\tilde{a}$ take a less familiar form, involving now generalized hypergeometric functions, as can be seen in eqs.~(\ref{15.03}) and (\ref{23}). This difference becomes more apparent when analyzing the uncertainty relation for the nonlinear coherent states: it has a minimum value, equal to $1/2$, for the nonlinear coherent states of eq.~(\ref{15.03}) and it reaches a maximum value equal to $3/2$ for the states of eq. (\ref{23}) (see figs. \ref{fig:nl} and \ref{fig:nl1}).
	
	On the other hand, the study of the eigenstates of the SAO for the SUSY harmonic oscillator involves naturally generalizations of the standard coherent states \cite{aragone,hussin,zypman}. As mentioned first in this paper, the analysis of the linear supercoherent states rests on the study of the matrix $K$, given in eq. (\ref{40.41}), its matrix elements $k_i$ and the behavior of its eigenvalues $\kappa_{\pm}$. Thus, a general expression for the eigenvectors $|Z\rangle$ and their Heisenberg uncertainty relation were found in \cite{zypman}.
	
	Here we have used the same treatment to find the nonlinear supercoherent states associated to the deformed SAO of the system, which instead of involving the standard coherent states associated to $\hat{a}$ depend now on the nonlinear ones for $\tilde{a}$. Moreover, the Heisenberg uncertainty relation for the nonlinear supercoherent states shows a different behavior as compared with the corresponding relation for the linear ones. This is so since the mean values of the operators $\hat{x}$, $\hat{p}$ and their squares become quite different for the several coherent states we are dealing with, either linear or nonlinear.
	
	As it was mentioned in subsection \ref{subsec:level22}, the deformations we are studying are also connected with the well known $f$-oscillators which, in turn, are extensions of the $q$-deformations of the oscillator. For these systems it turns out that the frequency depends on the amplitude through an arbitrary function \cite{ammsz00}. This property is not shared by our SUSY harmonic oscillator since the associated Hamiltonian is independent of the $f$-deformation, which does not happen for the nonlinear $f$-oscillators.
	
	Let us remark that, although $[\hat{H},\hat{A}]=-\omega\hat{A}$ and $[\hat{H},\hat{A}^\dagger]=\omega\hat{A}^\dagger$, the set $\{\hat{H},\hat{A}^\dagger,\hat{A}\}$ neither generates a Lie algebra nor a polynomial Heisenberg algebra \cite{fh99,cfnn04}, since $[\hat{A},\hat{A}^\dagger]$ neither is a linear combination of $\hat{H}$, $\hat{A}^\dagger$, $\hat{A}$ nor is a polynomial in $\hat{H}$. It remains open to determine to which algebraic structure the generators $\hat{H}$, $\hat{A}^\dagger$, $\hat{A}$ give place.
	
	Finally, according to previous analysis for systems with equidistant spectrum \cite{djf94}, and taking into account that the SUSY harmonic oscillator has precisely this property, it has been possible to determine both the system evolution loop as well as the geometric phases $\beta$ associated to the nonlinear supercoherent states of eqs.~(\ref{52.5}) and (\ref{68}).
	
	\section*{Acknowledgments}
	The authors acknowledge the support of Conacyt. EDB also acknowledges the Conacyt fellowship 290649.
	%
	%
	%

	%
	%
	
\end{document}